\documentclass[preprint,tightenlines,noshowpacs,nofootinbib,amsmath,amssymb]{revtex4-1}

\usepackage{amssymb}
\usepackage{amsmath}
\usepackage{amsfonts}
\usepackage{graphicx, float}
\newcommand\leftidx[3]{%
  {\vphantom{#2}}#1#2#3%
}

\begin{document}

\title{Connecting horizon pixels and interior voxels of a black hole}

\author{Piero~Nicolini}
\email{nicolini@fias.uni-frankfurt.de} 
\affiliation{Frankfurt Institute for Advanced Studies (FIAS), Ruth-Moufang-Str. 1, 
60438 Frankfurt am Main, Germany\\  \& \\ Institut f\"{u}r Theoretische Physik,
%Johann Wolfgang 
J.W. Goethe-Universit\"{a}t, Max-von-Laue-Str. 1,
60438 Frankfurt am Main, Germany
}

\author{Douglas Singleton}
\email{dougs@csufresno.edu}

\affiliation{Department of Physics, California State University, Fresno, CA 93740-8031 USA \\
\&\\
Instituto de Ciencias Nucleares, Universidad Nacional Aut{\'o}noma de M{\'e}xico,
Apartado Postal 70-543, Distrito Federal, 04510, M{\'e}xico}

\date{\today }

\begin{abstract}
In this paper we discuss to what extent one can infer details of the interior structure of a 
black hole based on its horizon. Recalling that black hole thermal properties are connected to the non-classical 
nature of gravity, we circumvent the restrictions of the no hair theorem by postulating  that the black hole 
interior is singularity free due to violations of the usual energy conditions. Further these conditions allow 
one to establish a one-to-one, holographic projection between Planckian areal ``bits'' on the horizon 
and ``voxels'', representing the  gravitational degrees of freedom in the black hole interior.
We illustrate the repercussions of this idea by discussing an example of the black hole interior 
consisting of a de Sitter core  postulated to arise from the local graviton quantum vacuum energy. 
It is shown that the black hole entropy can emerge as  the statistical entropy of a gas of voxels.
\end{abstract}

\maketitle

Classically, black holes hide their interior behind an event horizon. Since not even light can escape from 
behind the event horizon one can not (apparently) learn much about the interior structure of the black hole. 
According to the ``no-hair" theorem \cite{no-hair} the only thing one can learn about a black hole from outside 
the horizon is its mass, charge and angular momentum. Here ``hair" means any qualities other than mass, charge or 
angular momentum ({\it e.g.} baryon number) which characterizes the matter that formed the black hole. 
However, combining quantum field theory with black holes one finds black holes are not entirely black -- they
emit thermal radiation at the Hawking temperature $T_\mathrm{H}$ \cite{hawking}. Since black holes have a
temperature one can consider the possibility that they have an entropy. If we assume that 
 each fundamental particle of mass $m$ carries a basic unit of information, \textit{i.e.}, a bit, the 
total information associated with an imploding star of mass $M$ is roughly $M/m$. In turn these particles
of mass $m$ can fit into the black hole, provided that their Compton wavelength does not exceed the hole size, 
\textit{i.e.}, $\hbar/m c\lesssim r_\mathrm{S}$ with $r_\mathrm{S} = 2 G M /c^2$ being the horizon radius. 
By combining these conditions one can assign a black hole an entropy proportional to its horizon surface area 
\cite{bekenstein}
\begin{equation}
\label{entropy}
S_\mathrm{BH}  \sim \frac{k_\mathrm{B}G M^2}{\hbar c} \propto \frac{k_\mathrm{B} A}{l_\mathrm{Pl} ^2} ~,
\end{equation}
where $k_\mathrm{B}$ is Boltzmann's constant, $l_\mathrm{Pl} = \sqrt{G \hbar / c^3} \approx 10^{-35}$ m is 
the Planck length, and $A$ is the surface area of the black hole's horizon given by $A=4 \pi r_{\rm S} ^2$. 
\footnote{As an historical inversion the idea that black holes had an entropy was proposed 
first \cite{bekenstein} and after the precise formula for the temperature was given by Hawking 
\cite{hawking} the exact proportionality between the entropy and surface area was determined, giving
the well-known result $S_\mathrm{BH} = \frac{k_\mathrm{B} A}{4l_\mathrm{Pl} ^2}$.} 
One possible way to interpret equation \eqref{entropy} is that the surface area of the black hole, $A$, can be 
broken up into fundamental units or ``bits" of Planckian area $l_\mathrm{Pl} ^2$. Thus the horizon area $A$ can be 
seen as being tiled by Planckian area plaquettes. Since the Planck area is very small (on the order of 
$10 ^{-70} \ \mathrm{m}^2$) and since $A$ for stellar mass black holes will be, at minimum, 
of the order of $10 ^3 \ \mathrm{km} ^2$, one can see that the entropy associated with astrophysical 
black holes will be huge.  This can be seen by comparing the entropy of a solar mass black hole with
the entropy of a ``star" of photons with a temperature $T = 1000 ^o {\rm K}$. This temperature 
gives a photon wavelength of $\lambda \sim 10^{-7} {\rm m}$. The number of photons in such a ``star" is
$N_\mathrm{\gamma} \sim V/\lambda ^3 \sim (R_{\rm s}/\lambda )^3$ where $R_{\rm s}$ is the stellar radius. The entropy of this photon
star (normalized by $k_B$) is proportional to the number of photons 
$S_\mathrm{\gamma}/k_B \propto N_\mathrm{\gamma} \sim (R_{\rm s}/\lambda )^3$. Now if we take 
$R_{\rm s}$ to be of the order of the radius of the Sun, $R_{\rm s}\sim 10^9$ m, we get  $S_\mathrm{\gamma} / k_B \sim 10^{48}$.
In comparison the entropy for a solar mass black hole (again normalized by $k_B$) 
given by \eqref{entropy} (using $M  \sim  10^{30}$ kg) 
is $S_\mathrm{BH} / k_B \sim 10^{75}$ i.e. $S_\mathrm{BH} \gg S_\mathrm{\gamma}$.

Aside from a few special cases \cite{sv96,carlo,ash} the nature of the black hole entropy is not 
completely understood. Other than the thermodynamic definition $S_\mathrm{BH}\equiv\int \mathrm{d}M/T_\mathrm{H}$, 
we still lack a satisfactory statistical description of black hole entropy in terms of microscopic 
degrees of freedom. This can be re-phrased by saying one generally ignores the connection between 
the information/entropy of the horizon as seen by an external observer and the unknown, microscopic 
gravitational degrees of freedom in the interior of the black hole. In this paper we explore 
some concrete example of a one-to-one connection between the horizon 
entropy and some yet unspecified ``bits" or units in the black hole interior.

The fact that the entropy of a black hole scales as an area rather than as a volume is unusual. 
Normally the entropy of a system scales like the volume of the system.  This feature has led to 
the connection between gravitational systems and holography. 
In a black hole all the information/entropy is apparently encoded in the two dimensional horizon. 
This holographic approach to gravity was first expounded in \cite{thooft,susskind}, and an
overview of this subject can be found in \cite{susskind2}.  
We are proposing, that as with a real hologram, there should 
be some holographic projection between the  areal``bits" which tile the horizon 
and some ``voxels"\footnote{The name is a combination  of ``volume" and ``pixel". } or volume 
``bits" of the interior.  Voxels in general mean a three dimensional ``bit", but here we will have a 
generalized meaning of voxel as an ($n-1$)-spatial-dimensional volume ``bit". The reason for this is that in 
the interior of a black hole one has a fantastic high energy/high density natural laboratory. As one approaches the 
center of a Schwarzschild black hole one reaches energies and densities which are unattainable in a man-made 
laboratory. In fact for a classical Schwarzschild black hole the energy density diverges at the central
singularity. This singularity is problematic and probably signals the breakdown of general relativity.   

The main question is the character of these interior voxels. This is a fundamental issue of quantum gravity. 
In a nutshell one can say that quantum gravity plays the analogue of a kinetic theory of voxels, since it is 
expected to connect microscopic degrees of freedom to the thermodynamic value of the entropy in \eqref{entropy} 
(see for instance \cite{tj95,tanu02,tanu10,ev11,pn10}). As a result the voxels should be related to 
some short scale modification of the gravitational field. By assuming some condition which
leads to the avoidance of the interior singularities one can obtain possible ultraviolet completions 
of black hole space-times.  Due to the character of black hole metric 
coefficients, \textit{i.e.}, $g_{00}=(1+2\Phi_\mathrm{N}(r))$, the Newtonian potential $\Phi_\mathrm{N}$ 
provides enough information for this purpose -- static gravitational forces are the result of virtual graviton exchange that 
can be described by  a scalar theory. Writing the Newtonian potential as
\begin{equation}
 \Phi_\mathrm{N}(r) = \frac{M}{M_\mathrm{Pl}^2}\int \frac{\mathrm{d}^3k}{(2 \pi )^3}  \; \; 
D\left ( k \right )\mid_{k_0=0} \; \exp\left ( i \vec k \cdot  \vec r   \right ) 
\end{equation}
one wants ultraviolet finite propagators, $D(k)$, to tame the classical curvature singularity. 
The specific way the propagator is modified corresponds to specific models of a quantum gravity improved black hole.

Despite the different approaches to quantum gravity, the regularity of space-time requires that 
metric coefficients fulfill the condition $|\partial_r^2 g_{00}|<M_\mathrm{pl}^2$, which is equivalent 
to saying that the curvature can at most assume Planckian values. The above condition is easily met 
for distances $\gg l_\mathrm{Pl}$. On the other hand, at scales $\sim l_\mathrm{Pl}$, one ends up with a 
Netwonian potential $\Phi_\mathrm{N}(r)\sim {\cal O}(r^{2})$. According to this reasoning, a simple 
realization of a regular space-time is based on the assumption that the center of the black hole is replaced 
by a de Sitter core \cite{bardeen}. The latter has been considered in early attempts to avoid 
the curvature singularity by matching an outer Schwarzschild geometry with an inner de Sitter geometry along 
time-like \cite{aa1,aa2,aa3} and space-like matter \cite{fmm} shells.  Physically, a
de Sitter core is a repulsive gravity region which can prevent the complete gravitational collapse to a
singular matter/energy density profile. Local violations of energy conditions are 
the signature for the non-classical nature of the resulting black hole at short scales.
There are a host of different approaches \cite{dym, dym2, nss06,ns10,pn09, mmn11,pn12, hay, stefano} to avoiding 
the central singularity which all amount to having a de Sitter or de Sitter-like core inside the horizon.

At present a consistent quantum theory of matter and gravity in the interior of a black hole does not exist.
Thus we will keep our picture of the interior as general as possible. The first assumption mentioned above
is a generic de Sitter core. Secondly we allow the dimensionality of the de Sitter core to be larger or smaller than
four (3 space plus one time). We first discuss the possibility of a de Sitter core with space-time dimensions
four or larger. String theory, brane models, and Kaluza-Klein theories, are examples of theories 
allowing space-time dimensionality greater than four. The extra dimensions are `` curled-up" to a small
size/large energy scale so that one is not able to access these extra spatial dimensions except at high energy densities.
Although such energies may not be feasible in the laboratory, they can be reached at some point in the interior 
of a black hole, resulting in the extra dimensions ``opening up". Thus we model the interior of the black 
hole as an $n$-dimensional de Sitter space-time where at first we take $n \ge 4$. The interior metric is 
\begin{equation}
\label{dS}   
\mathrm{d}s _\mathrm{interior} ^2 = -\left( 1 -\frac{r^2}{\alpha^2} \right) \mathrm{d}t^2 
+ \frac{\mathrm{d}r^2}{\left( 1 -\frac{r^2}{\alpha^2} \right)} +r^2 \mathrm{d} \Omega _{n-2} ~.  
\end{equation}
$d \Omega_{n-2}$ is the differential angular part of the metric for the angular coordinates $\theta, \phi_i$ where
$i=1, 2, 3, ..., n-3$. The constant $\alpha$ is related to the positive cosmological constant $\Lambda$ by 
$\Lambda (n) = \frac{(n-2)(n-1)}{2 \alpha ^2}$. For the usual case where $n=4$ this leads to  
$\Lambda = \frac{3}{ \alpha ^2}$.

For the exterior metric we assume (for simplicity) a Schwarzschild metric 
\begin{equation}
\label{schwarz}   
\mathrm{d}s _\mathrm{exterior} ^2 = -\left( 1 -\frac{2GM}{c^2 r} \right) \mathrm{d}t^2 + 
\frac{\mathrm{d}r^2}{\left( 1 -\frac{2GM}{c^2 r} \right)}
+r^2 \mathrm{d} \Omega _2 ~.  
\end{equation}  
%In this paper we are not concerned about the details of the transition between the above two geometries. 
To draw our conclusions we only need the asymptotic forms of the above metrics.
We offer, however, an example to clarify how the smooth transition between the metrics with different dimensions can take place. 
We follow the standard arguments of terascale black holes with special reference to the large extra dimension paradigm. 
According to ADD proposal \cite{ADD,ADD2}, 
the additional spatial dimensions must have a size $R$ small enough to be usually un-observable: Gravity has to have the standard behavior at macroscopic scales. On the contrary for distances $r\lesssim R$, gravity can probe the full $(4+d)$-dimensional \textit{bulk} spacetime ${\cal M}^{(4+d)}$ with $n=4+d$. The bulk can be factorized as ${\cal M}^{(4+d)}={\cal M}^{(4)}\times T^{(d)}$, where  ${\cal M}^{(4)}$ is the \textit{brane} (\textit{i.e.} our standard four dimensional Universe) and $T^{(d)}$ is a $d$-dimensional torus with radii of size $R$. Einstein's equations can be derived from the action $S=S_\mathrm{g}+S_\mathrm{m}$, where the gravitational part reads
\begin{eqnarray}
&&S_\mathrm{g}\sim M_\mathrm{F}^{d+2}\int {\cal R}\sqrt{-g}\ \mathrm{d}^{n} x ,
\end{eqnarray}
where $M_\mathrm{F}$ is the higher dimensional, fundamental scale. By performing a Kaluza-Klein (KK) expansion of the graviton field one can obtain a dimensionally reduced action which is valid for $r\gg R$.
\begin{eqnarray}
&&S_\mathrm{g}\to  \underbrace{M_\mathrm{F}^{d+2}R^d \int \leftidx{^{(4)}}{{\cal R}}{}\sqrt{-\leftidx{^{(4)}}{g}{}}\ \mathrm{d}^4 x}_{\mathrm{effective\ brane\ action}} +\underbrace{\sum_{k>0}\left(\dots\right)}_{\mathrm{KK\ excitations}}.
\end{eqnarray}
This is another way to saying that, up to sub-leading corrections, gravity behaves normally at macroscopic scales, \textit{i.e.} as Einstein gravity  \cite{Shifman,BN14}. The matter action, $S_\mathrm{m}$, contains  a term depending on Standard Model (SM) fields that do not propagate in the bulk
\begin{equation}
 \int {\cal L}_{\rm SM}(\Phi_{\rm SM}) \ \sqrt{-\leftidx{^{(4)}}{g}{}} \ \mathrm{d}^4x .
\end{equation}
Since we are concerned about neutral black hole solutions we will not consider this term.
In the regime $r< R$, however, there has to be a mass density that generates the $(4+d)$-dimensional de Sitter core. 
This means that $S_\mathrm{m}$ has an additional term of the form
\begin{equation}
\int {\cal L}_{\rm dS}\ \sqrt{-g} \ \mathrm{d}^nx .  
\end{equation}
From the term above one obtains an energy-momentum tensor $T^{\mu\nu}_{\rm dS}$ that must vanish outside the bulk. With the above ingredients one can actually find black hole solutions with radii $r_\mathrm{S}= 2GM/c^2\gg R$ that can be described by the $4$-dimensional line element $\mathrm{d}s_\mathrm{exterior}^2$. At distances smaller than $R$, however, the geometry behaves drastically different from the usual black hole solution since a de Sitter core, described by the line element $\mathrm{d}s_\mathrm{interior}^2$,  forms in the bulk. Apart from the out horizon $r_\mathrm{S} = 2GM/c^2$, there exist an inner horizon in the bulk, located at  $r_\mathrm{dS} = \alpha = \sqrt{\frac{(n-2)(n-1)}{2 \Lambda}}$. It turns to be that $r_\mathrm{dS} \ll r_\mathrm{S}$.

The manner in which we connect the horizon with the interior is simply by the requirement that every area
plaquette on the horizon should have some corresponding $(n-1)$-spatial-dimensional ``voxel" in the interior. 
The spatial volume $V_{n-1}$ associated with the $n$-dimensional de Sitter metric \eqref{dS} is
\begin{equation}
\label{vol}
V_{n-1} = \int _0 ^\alpha \frac{S_{n-2} r^{n-2}}{\sqrt{1-\frac{r^2}{\alpha ^2}}} \mathrm{d}r = 
\frac{\pi ^{n/2} \alpha ^{n-1}}{\Gamma(n/2)} ~,
\end{equation}
where $\Gamma (x)$ is the Gamma-function and $S_{n-2} = \frac{2 \pi ^{(n-1)/2}}{\Gamma((n-1)/2)}$
is the surface area of the unit $(n-2)$-sphere. Now if we take the length scale of the fundamental 
$(n-1)$-spatial-dimensional voxel that builds up $V_{n-1}$ to be $l$, then the total number of voxels in 
$V_{n-1}$ is
\begin{equation}
\label{nvoxel}
N_\mathrm{voxel}\sim\frac{V_{n-1}}{l^{n-1}} = \frac{\pi ^{n/2}}{\Gamma(n/2)} \left( \frac{\alpha}{l} \right)^{n-1}~.
\end{equation}
Note that we have left the scale of the fundamental voxel length (\textit{i.e.} $l$) arbitrary. 
One might assume this should be of the order of the Planck length (\textit{i.e.} $l \simeq l_\mathrm{Pl}$), 
or some other higher dimensional fundamental scale $l_\ast$ (\textit{i.e.}  $l \simeq l_\ast$)  
or maybe it would be set of the scale $\alpha$ (i.e. $l \simeq \alpha$). However, here we leave $l$ 
free and not necessarily connected with the other scales of the system. The assumption of this paper 
is that, aside from numerical factors 
involving $\pi$ or $\Gamma(x)$, the number of horizon bits/pixels (\textit{i.e.} 
$N_\mathrm{pixel}\sim A/l_\mathrm{Pl}^2 \sim r_\mathrm{S} ^2/ l ^2 _\mathrm{Pl}$) should coincide 
with the interior voxels (\textit{i.e.} $N_\mathrm{voxel}\sim V_n/l^{n-1} \sim (\alpha / l)^{n-1}$). Explicitly this gives  
\begin{equation}
\label{info-cons}
N_\mathrm{voxel}=N_\mathrm{pixel}\Rightarrow\frac{\alpha}{l} \sim \left(\frac{r_\mathrm{S}}{l_\mathrm{Pl}} \right)^{2/(n-1)}~.
\end{equation}
Now taking $r_\mathrm{S}$ to be the Schwarzschild radius of some astrophysical black hole so that 
$r_\mathrm{S} \approx 10 ^4$ m one can see that the ratio of the right hand side of \eqref{info-cons} is of 
the order $r_\mathrm{S} / l_\mathrm{Pl} \approx 10^{39}$ -- very large. For the case of $4$-dimensional
space-time ({\it i.e.} when $n=4$) one finds $\alpha/l\sim 10^{26}$. In such a case we do not have any constraint on the size of the de Sitter core as in the presence of extra dimensions. We just require that $r_\mathrm{dS} \le r_\mathrm{S}$, so the de Sitter core is inside the Schwarzschild horizon.  For $\alpha=10^3$ m (which is a distance 
scale still less than the horizon radius of stellar black holes) this gives $l\sim 10^{-23}$ m, 
which is a distance scale that can potentially be probed. In such a scenario one does not necessarily need to 
invoke extra-dimensions to get observable results. The effect of considering extra dimensions is that by increasing 
the number $n$, the ratio $\alpha/l$ decreases. We need, however, to take into account that also the size of extra dimensions decreases with $n$. For $M_ \mathrm{F} \sim 1$ TeV, $R$ becomes smaller like $\sim 10^{\frac{32}{d} - 19}$ meters as $d$ increases \cite{PDG,BN14}. By recalling that $\alpha\lesssim R$, we find that $l$ is nearly stable versus $n$, \textit{i.e.}, the value of $l$ is  in the range 
$10^{-22}-10^{-23}$ m. Table \ref{tab} summarizes the main quantities as $n$ varies. We consider $n\geq 7$ only,  since $n=5, 6$ are experimentally ruled out \cite{BN14}.
\begin{table}[htbp]
\caption{Values and constraints on the parameters, $R$, $\alpha$, $l$ for different space-time dimensions $n$ for
$M_F \sim 1 $ TeV}
\begin{tabular}{|c|llll|}
 \hline
$n$ & $\phantom{1.0\ }7$ & $\phantom{1.4\ }8$ & $\phantom{5.6\ }9$ & $\phantom{4.6 }10$ \\ \hline\hline
$\alpha/l\sim$ & $1.0\times 10^{13}$  & $1.4\times 10^{11}$  & $5.6\times 10^{9}$ & $4.6\times 10^{8}$   \\
  $R[{\rm m}]\sim$  & $4.6\times 10^{-9}$ & $1.0\times 10^{-11}$  & $2.5\times 10^{-13}$ & $2.2\times 10^{-14}$ \\
$l[{\rm m}]\lesssim$ & $4.6\times 10^{-22}$  & $7.1\times 10^{-23}$ & $4.5\times 10^{-23}$ &$4.8\times 10^{-23}$ \\ \hline
\end{tabular}
\label{tab}
\end{table}
The result is that the fundamental length scale, $l$, in 
the black hole interior can be different from both $l_\mathrm{Pl}$ and $\alpha$.

Note that our conjecture \eqref{info-cons}, supported by a simple argument of information conservation, 
can be further improved by requiring that the black hole entropy is nothing but the entropy of a ``gas of voxels'', namely
\begin{equation}
S_\mathrm{voxel}\equiv k_\mathrm{B}\log \left((N_\mathrm{voxel})!\; / \; \prod_i N_i! \right)\stackrel{!}{=} S_\mathrm{BH} ~,
\end{equation}
where $N_i$ are voxel-admissible microscopic configurations. In the limit of large $N_\mathrm{voxel}$ this gives
\begin{equation}
 N_\mathrm{voxel}\sim N_\mathrm{pixel}.
\end{equation}
Apart from logarithmic corrections associated with the probability of each voxel configuration, the above condition 
offers a way to interpret, at a statistical level, the black hole entropy in terms of an $(n-1)$-spatial-volume 
dependent entropy. Thus the areal dependence of $S_\mathrm{BH}$ would 
be a thermodynamic fictitious effect seen by an asymptotic observer.

We now turn to the opposite possibility: that the number of dimensions of the de Sitter
core may be less than four. This reduced dimensional scenario is called spontaneous dimensional reduction 
\cite{thooft,carlip1,carlip2}. The basic idea is that the spacetime in its high energy/short distance regime might switch 
from the conventional differential manifold configuration to that of a fractal surface as a result of a huge loss of 
local resolution. As a result, the actual space-time dimension would be expressed in terms of some fractal dimension, 
able to smoothly ``flow" from the conventional topological value, four, to some smaller value, \textit{e.g.}, two. 
The idea that space-time dimensionality can vary with energy is nowadays supported by an array of 
quantum gravity models and numerical experiments for fractal space-times 
\cite{amb,reu,dario,leonardo,lmpn,pnes,lattice,gianluca1,gianluca2}. As  noted in \cite{mn13} this has repercussions 
for black hole metrics: Space-times are conformal invariant; it is no longer possible to distinguish between 
small/big or classical/quantum black holes \cite{robb,daniel1,daniel2}; the two dimensional gravitational coupling, $G_2$, 
is dimensionless. These properties are already evident by examining the Newtonian potential for a
mass $M$ in $(1 + 1)$-dimensions,
{\it i.e.},
$$
\Phi _2 \simeq G_2 M x ~,
$$
which in unaffected by any rescaling of lengths $x$ and masses $M$, as in the case
of the quantum mechanical Compton relation governing particle sizes. This is
confirmed by full metric of dilaton gravity black holes, {\it e.g.} \cite{robb} 
\begin{equation}
\label{dilaton-grav}
ds_2 ^2 = - \left( \frac{2 G_2 M}{c^2} |x| + C \right) dt^2 + \frac{dx^2}{\left( \frac{2 G_2 M}{c^2} |x| + C \right)} ~,
\end{equation}
whose gravitational radius is proportional to the inverse of the mass, $r_S \simeq 1/M$.
As a consequence, in 2 dimensions, we no longer have a minimal length
scale, encoded in the gravitational coupling, at which further compression of
particles is prevented by matter collapsing into a black hole. Nevertheless it has been
shown, that as in the higher dimensional case, regularizing quantum effects can replace 
the singularity with a de Sitter core in two dimensions \cite{dea,mn11}.

To obtain spacetimes in both the large and short scale regimes, we need an action for the dimensionally reduced phase. Among the class of dilaton gravity models, the following action can be obtained from the Einstein-Hilbert action in the limit $n\to 2$ \cite{robb} (see also other dimensionally reduced models in \cite{jack,daniel1,daniel2})
\begin{equation}
S_{2} = \int d^2x~\sqrt{-\leftidx{^{(2)}}{g}{}} \left[\left(\frac{c^4}{8\pi G_1} \psi\ \leftidx{^{(2)}}{{\cal R}}{}-\frac{1}{2}(\nabla \psi)^2\right) +{\cal L}_{\rm m}^{(1+1)}\right].
\end{equation} 
Here $\Psi$ is the dilaton field and ${\cal L}_{\rm m}^{(1+1)}$
is the dimensionally reduced matter Lagrangian. The last step is to assume a suitable ${\cal L}_{\rm m}^{(1+1)}$  describing the de Sitter core as a solution of equations derived from the above action. In such a way the black hole will look $4$-dimensional to an external observer but will also exhibit a regular $2$-dimensional interior.

As a result we assume that our inner metric is described by the short 
scale behavior of a regular dilaton gravity black hole given by \cite{dea}
\begin{equation}
\label{dS2}   
\mathrm{d}s _\mathrm{interior} ^2 = -\left( \frac{r^2}{\alpha^2}-C \right) \mathrm{d}t^2 + 
\frac{\mathrm{d}r^2}{\left( \frac{r^2}{\alpha^2} -C\right)}
\end{equation}
where $C$ is a free parameter. The space-time structure is quite different from the higher-dimensional counterparts. 
Regardless of the sign of $C$ the space-time is well behaved at $r=0$ and one can obtain the spatial volume of the core by
\begin{equation}
\label{vol-2}
V_1 = \int _0 ^{\alpha \sqrt{|C|}} \frac{S_{0}}{\sqrt{\left|\frac{r^2}{\alpha ^2} -C\right|}} \mathrm{d}r\sim \alpha ~,
\end{equation}
\textit{i.e.} the core volume is of order $\alpha$. Accordingly, the number of voxels is $N_\mathrm{voxel}\sim \alpha/l$. 
In two dimensions, since $G_2$ is dimensionless, there is no minimum length scale, thus we can adjust the parameter $l$  
to any arbitrarily small value  in order to match the number of area bits $N_\mathrm{pixel}$ on the horizon. 

The main idea of this paper is that the horizon can be connected to the internal structure of black holes.
Our line of reasoning is based on the existence of a relation between the areal ``bits"/plaquettes on the horizon and some  
$(n-1)$-spatial-dimensional voxels of the interior of the black hole. 
To frame our arguments we modeled the black hole interior by means of an $(n)$-dimensional de Sitter space-time. 
This is a generic way for avoiding the central singularity, which is supported by several quantum gravity 
improved black hole models.  Thus the voxels, which compose the interior volume, are $(n-1)$-spatial-dimensional cubes 
of de Sitter space. By adjusting the dimensionality, $n$, of the de Sitter space and/or the de Sitter length scale 
$\alpha$ (which is the same as adjusting the cosmological constant for the vacuum 
inside the black holes since $\Lambda \propto \alpha ^{-2}$) one finds that the voxel length scale, $l$, in the interior
of the black hole can be very different from either the Planck length scale, $l_\mathrm{Pl}$, or the cosmological length 
scale, $\alpha$. We showed that this kind of reasoning can offer an interpretation of the areal entropy of black holes 
in terms of a volume depending statistical entropy associated to a gas of voxels. 
We also considered the case of spontaneous dimensional reduction of black hole interiors.  
We showed that the parameter freedom of dilaton gravity black holes can be exploited in order to match internal 
$1$-dimensional voxels with the horizon bits. 

Despite the provisional nature of the proposed arguments, we believe 
that the concept of fundamental voxel encoding gravitational information could lead to new insights about the nature 
of black hole entropy. \\

{\it Note Added:} After this paper was accepted we learned of two works \cite{jabbari} \cite{rama} which discuss
some of the issues dealt with in this paper.

\begin{acknowledgements}
This work has been supported by the grant NI 1282/3-1 of the project ``Evaporation of microscopic black holes" of 
the German Research Foundation (DFG), by the Helmholtz International Center for FAIR within the framework of the 
LOEWE program (Landesoffensive zur Entwicklung Wissenschaftlich-\"{O}konomischer Exzellenz) launched by the State 
of Hesse and partially by the European Cooperation in Science and Technology (COST) action MP0905 ``Black Holes 
in a Violent Universe''. D.S. would like to thank the generous hospitality of the Frankfurt Institute for
Advanced Studies (FIAS), at which this work was initiated as well as Universidad Nacional Aut{\'o}noma de M{\'e}xico where
part of this work was completed. D.S. would also acknowledges the support of an APS ITGAP grant.
\end{acknowledgements}

\end{document}